\documentclass[aip,apl,reprint]{revtex4-2}

\usepackage{epsfig,graphicx,times}
\usepackage{bm}
\usepackage[T1]{fontenc}
\usepackage{mathptmx}
\usepackage{etoolbox}
\usepackage{ulem}
\usepackage{color}
\usepackage{amstext}
\usepackage{amsmath}            
\usepackage{amssymb}            
\usepackage{graphicx}           
\usepackage{latexsym}

\begin{document}

\title{Broadband merged-element Josephson parametric amplifier}

\author{Yuting Sun}
\email[]{qu_syt@163.com}
\affiliation{National Laboratory of Solid State Microstructures, School of Physics, Nanjing University, Nanjing 210093, China}
\affiliation{Shishan Laboratory, Suzhou Campus of Nanjing University, Suzhou 215163, China}
\affiliation{Hefei National Laboratory, Hefei 230088, China}
\author{Xianke Li}
\affiliation{National Laboratory of Solid State Microstructures, School of Physics, Nanjing University, Nanjing 210093, China}
\affiliation{Shishan Laboratory, Suzhou Campus of Nanjing University, Suzhou 215163, China}
\affiliation{Hefei National Laboratory, Hefei 230088, China}
\author{Qingyu Wang}
\affiliation{National Laboratory of Solid State Microstructures, School of Physics, Nanjing University, Nanjing 210093, China}
\affiliation{Shishan Laboratory, Suzhou Campus of Nanjing University, Suzhou 215163, China}
\author{Tairong Bai}
\affiliation{National Laboratory of Solid State Microstructures, School of Physics, Nanjing University, Nanjing 210093, China}
\affiliation{Shishan Laboratory, Suzhou Campus of Nanjing University, Suzhou 215163, China}
\affiliation{Hefei National Laboratory, Hefei 230088, China}
\author{Xudong Liao}
\affiliation{National Laboratory of Solid State Microstructures, School of Physics, Nanjing University, Nanjing 210093, China}
\affiliation{Shishan Laboratory, Suzhou Campus of Nanjing University, Suzhou 215163, China}
\author{Dong Lan}
\affiliation{National Laboratory of Solid State Microstructures, School of Physics, Nanjing University, Nanjing 210093, China}
\affiliation{Shishan Laboratory, Suzhou Campus of Nanjing University, Suzhou 215163, China}
\affiliation{Hefei National Laboratory, Hefei 230088, China}
\affiliation{Synergetic Innovation Center of Quantum Information and Quantum Physics, University of Science and Technology of China, Hefei, Anhui 230026, China}
\author{Jie Zhao}
\email[]{jiezhao@nju.edu.cn}
\affiliation{National Laboratory of Solid State Microstructures, School of Physics, Nanjing University, Nanjing 210093, China}
\affiliation{Shishan Laboratory, Suzhou Campus of Nanjing University, Suzhou 215163, China}
\affiliation{Hefei National Laboratory, Hefei 230088, China}
\author{Yang Yu}
\affiliation{National Laboratory of Solid State Microstructures, School of Physics, Nanjing University, Nanjing 210093, China}
\affiliation{Shishan Laboratory, Suzhou Campus of Nanjing University, Suzhou 215163, China}
\affiliation{Hefei National Laboratory, Hefei 230088, China}
\affiliation{Synergetic Innovation Center of Quantum Information and Quantum Physics, University of Science and Technology of China, Hefei, Anhui 230026, China}

\date{June 6 2025}

\begin{abstract}
Broadband quantum-limited amplifiers are essential for quantum information processing, yet challenges in design and fabrication continue to hinder their widespread applications. Here, we introduce the broadband merged-element Josephson parametric amplifier in which the discrete parallel capacitor is directly integrated with the Josephson junctions. This merged-element design eliminates the shortcomings of discrete capacitors, simplifying the fabrication process, reducing the need for high-precision lithography tools, and ensuring compatibility with standard superconducting qubit fabrication procedures. Experimental results demonstrate a gain of 15 dB over a 500 MHz bandwidth, a mean saturation power of -116 dBm and near-quantum-limited noise performance.
This robust readily implemented parametric amplifier holds significant promise for broader applications in superconducting quantum information and the advancement of quantum computation.

\end{abstract}

\maketitle
Quantum limited parametric amplifiers are essential for quantum information and quantum metrology processes~\cite{readoutPRL2014,quantumsensor2017, darkmatterPRL, cosmic2013,darkmatterPRL,cosmic2013}. In superconducting quantum information, a quantum-limited Josephson parametric amplifier (JPA)~\cite{castellanos-beltranWidelyTunableParametric2007a,yamamotoFluxdrivenJosephsonParametric2008a,aumentadoSuperconductingParametricAmplifiers2020a} is typically used as the preamplifier to suppress the readout noise and enhance the signal-to-noise ratio (SNR). Josephson parametric amplifiers are key components for realizing high-fidelity single-shot readout~\cite{singleshot2009,singleshot2013}, generating squeezed quantum states~\cite{squeezedstatePRL2011_1,squeezedstatePRL2011_2}, implementing weak measurements~\cite{weak2013science,weak2013nature,weak2014nature}, performing real-time quantum feedback~\cite{feedback2012}, and achieving quantum error correction (QEC)~\cite{qec1,qec2,qec3}.

In the Noisy Intermediate-Scale Quantum (NISQ) era, a broadband parametric amplifier is necessary to fulfill the requirement of instantaneous readout of multiple qubits. Based on this requirement, a variety types of broadband Josephson parametric amplifiers have been proposed and experimentally demonstrated. 
They can be generally categorized into two types: the impedance-transformed parametric amplifier (IMPA)~\cite{hatridgeDispersiveMagnetometryQuantum2011a,mutusDesignCharacterizationLumped2013,mutusStrongEnvironmentalCoupling2014a,eloBroadbandLumpedelementJosephson2019,duanBroadbandFluxpumpedJosephson2021,grebelFluxpumpedImpedanceengineeredBroadband2021,luBroadbandJosephsonParametric2022,wuVacuumgapbasedLumpedElement2022,whiteReadoutQuantumProcessor2023,siddiqiPhysRevResearch,zhou2025impedance}, and the traveling wave parametric amplifier (TWPA)~\cite{macklinNearquantumlimitedJosephsonTravelingwave2015,obrienResonantPhaseMatching2014,bockstiegelDevelopmentBroadbandNbTiN2014,hoeomWidebandLownoiseSuperconducting2012,espositoPerspectiveTravelingWave2021}.
Although TWPAs are capable of achieving bandwidths greater than 1 GHz, their fabrication process is comparatively more complex. Furthermore, their gain profile is often insufficiently flat, and their ultra-wide bandwidth is frequently limited by other cryogenic microwave devices in the system. IMPAs are commonly used in superconducting quantum information experiments owing to their balance between gain, bandwidth, and fabrication compatibility~\cite{whiteReadoutQuantumProcessor2023}.

Lumped capacitors in parallel with the Josephson junctions play a crucial role in achieving simultaneously high gain and broad bandwidth in IMPAs. Various types of capacitors have been employed in IMPAs, such as interdigital capacitors~\cite{eloBroadbandLumpedelementJosephson2019}, vacuum gap capacitors~\cite{wuVacuumgapbasedLumpedElement2022}, and dielectric parallel-plate capacitors~\cite{mutusStrongEnvironmentalCoupling2014a, duanBroadbandFluxpumpedJosephson2021}. Each type of capacitor has its own shortcomings. The parasitic inductance of an interdigital capacitor is inevitable and limits the working frequency range of a parametric amplifier~\cite{wuVacuumgapbasedLumpedElement2022}. The yield of vacuum gap capacitors is hard to improve, and the stability of vacuum gap capacitors is generally inferior to that of other types capacitors. Dielectric parallel-plate capacitors normally require multi-layer fabrication processes, and imperfections in the dielectric layer can lead to defects and more dielectric losses. These issues introduce significant complexity and diminish the reliability of the IMPA fabrication process. The development of reliable and efficient IMPAs, particularly those compatible with superconducting qubit fabrication processes, continues to be a focus of research.

In this Letter, we introduce the merged-element technique to design and fabricate the so-called merged-element JPA (MEJPA). In our design, the self-capacitance inherent in the Josephson junctions replaces the discrete parallel capacitor, effectively merging the superconducting quantum interference device (SQUID) and the parallel capacitor into a single integrated element. Compared to the traditional IMPA designs, our approach streamlines the fabrication process by obviating the need for discrete parallel capacitors. Typical IMPAs require a SQUID with a critical current of several microamperes ($\rm \mu A$s) and a discrete capacitor with a capacitance of several picofarads ($\rm pF$s). In order to simultaneously meet the requirements for critical current and capacitance, Josephson junctions with junction areas on the order of tens of square micrometers were fabricated using the overlap method.~\cite{balOverlapJunctionsSuperconducting2021}. Owing to the concomitant influence of the Josephson junction oxidation layer on both critical current and junction capacitance, precise control of the oxidation parameters in the fabrication process is required. With the fabricated MEJPA, we achieved a gain of 15 dB (with 500 MHz bandwidth) and 10 dB (with 1 GHz bandwidth), operating near the quantum noise limit. Compared with the separate capacitor designs, merging the parallel capacitor into the Josephson junctions eliminates the shortcomings of conventional capacitors, makes the fabrication process fully compatible with standard superconducting qubit production procedures, and also streamlines the fabrication process. Furthermore, the use of large-area Josephson junctions circumvents the need for high-precision exposure tools, such as electron beam lithography (EBL), throughout the fabrication process.

\begin{figure*}[hptb]
\centering
\includegraphics[width=0.85\textwidth]{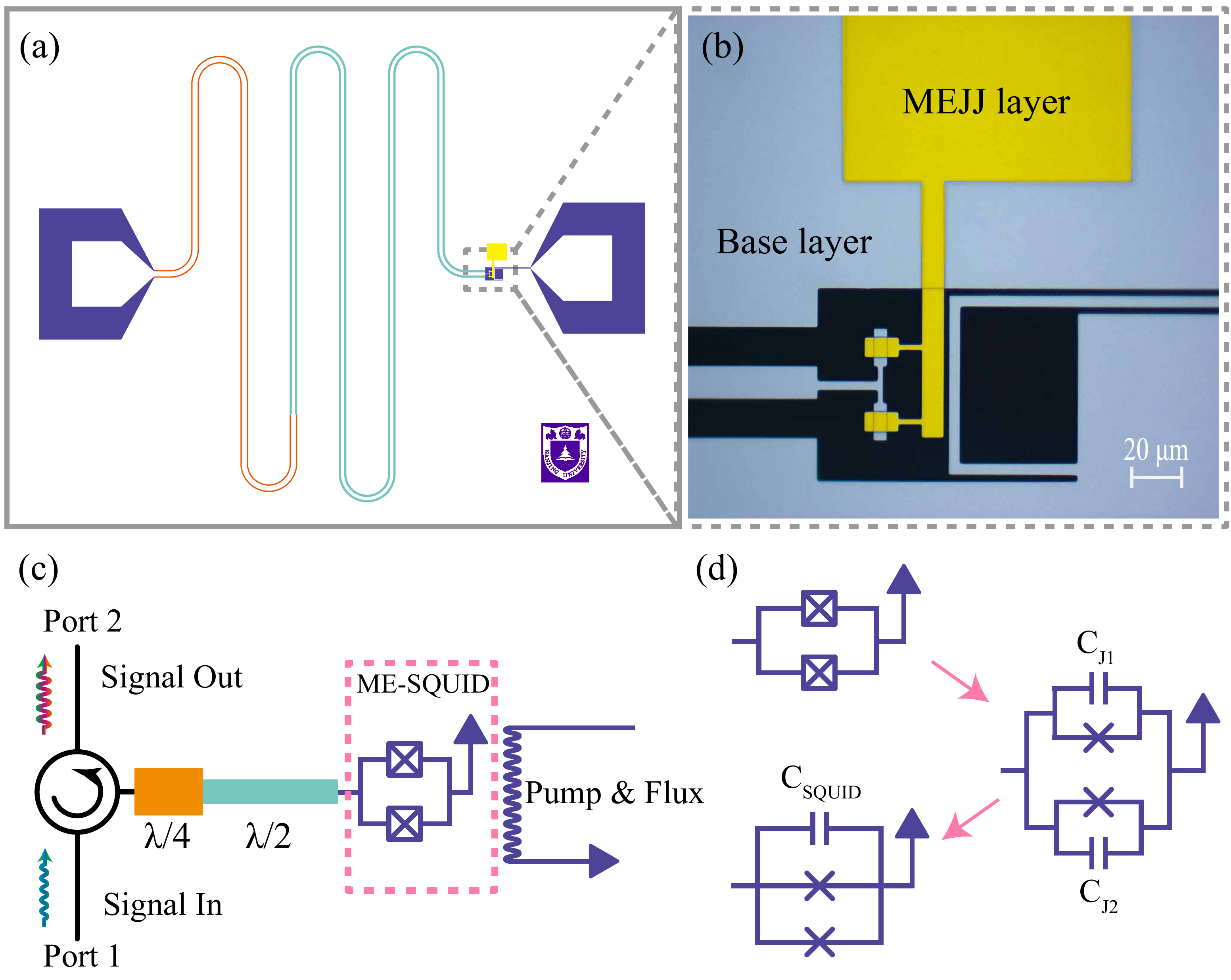}
\caption{\label{fig:fig1} (a) The layout of the merged-element broadband Josephson parametric amplifier. The CPW impedance transformer consists of a series-connected $\rm \lambda/4$ resonator (orange) and $\rm \lambda/2$ resonator (cyan), both designed to resonate at 6.5 GHz. (b) Optical micrograph of a zoomed-in view of the merged-element JPA. The entire chip consists of the base layer(dusty blue part) and the MEJJ layer(yellow part), which is highlighted in yellow false color. (c) The circuit diagram of our device. The left end of the $\rm \lambda/4$ resonator is wire-bonded to the sample box, and then connected to a circulator in the measurement setup via a short microwave cable. The right end of the $\rm \lambda/2$ resonator is terminated by the MEJPA device, which includes a flux-pumped DC SQUID connected to ground. (d) The equivalent models of the MEJPA, which is highlighted by the pink dashed box in (c).}
\end{figure*}

Figure.~\ref{fig:fig1}(a) shows the layout of an impedance-transformed MEJPA device. An impedance transformer is comprised of a series-connected $\rm \lambda/4$ coplanar waveguide (CPW) resonator with characteristic impedance $\rm Z_{\lambda/4}$ (the orange part) and a $\rm \lambda/2$ coplanar waveguide (CPW) resonator with characteristic impedance $\rm Z_{\lambda/2}$ (the cyan part). These two resonators are designed to operate at the same resonant frequency $\omega_t$. This impedance transformer is coupled to the MEJPA, enabling the effective impedance matching and optimizing the bandwidth performance. 
These two resonators are designed to operate at the same resonant frequency $\omega_t$= 6.5~GHz and are connected to the MEJPA, thereby enabling effective impedance matching and optimizing the bandwidth performance. The physical lengths of the $\rm \lambda/4$ and $\rm \lambda/2$ CPW sections on a sapphire substrate are approximately 4600 and 9200~$\mu$m, respectively. The left end of the $\rm \lambda/4$ resonator is wire-bonded to the sample box, and then connected to a circulator in the measurement setup via a short microwave cable, as shown in Fig.~\ref{fig:fig1}(c). The right end of the $\rm \lambda/2$ section is terminated by the MEJPA, which contains a flux-pumped DC SQUID. In our design, the shunt capacitor is merged into the SQUID, making the MEJPA appear as a DC SQUID. A detailed view of the DC SQUID and its flux-pump line is shown in Fig.~\ref{fig:fig1}(b). Figure.~\ref{fig:fig1}(c) illustrates the circuit diagram of the test setup for our device. The input signal goes through a circulator and is coupled to the MEJPA via an impedance transformer. The input impedance seen by the MEJPA is $\rm Z_{in}[\omega]$,
\begin{align}\label{eq1}\rm
    Z_{in}[\omega]&=R+\rm i\alpha \omega, \\
    \alpha &=\rm{2Z_{eff}}/\omega_{t}.
\end{align}

\noindent Here, $\rm Z_{eff}$ is the effective impedance of the impedance transformer, and the imaginary part $\rm \alpha$ can be tuned by the impedance transformer.~\cite{royBroadbandParametricAmplification2015,duanBroadbandFluxpumpedJosephson2021,PRXQuantum.3.020201} The input impedance $\rm Z_{in}[\omega]$ determines the reflection coefficient and thus the gain profile of the broadband parametric amplifier. The gain profile can be obtained by
\begin{align}\label{eq2}\rm
    G[\omega]&=G_0+G_2\omega^2+G_4\omega^4+\dots
\end{align}

\noindent Eliminating the quadratic coefficient $G_2$ by adjusting $\rm \alpha$ enhances the gain-bandwidth product. The resulting bandwidth, determined by the oscillator linewidth $\kappa_0$, is given by
\begin{align}\label{eq3}\rm
\Gamma_{BW}&=\frac{\kappa_0}{2}(\frac{1}{G_0})^{1/4}.
\end{align}

Typically, a DC SQUID primarily functions as a nonlinear inductor. In the MEJPA, the nonlinear inductor in conjunction with the inherent capacitance of the DC SQUID constitutes a lumped nonlinear oscillator. Therefore, the DC SQUID in our device performs the same function as a JPA with a discrete parallel capacitor. The equivalent circuit models of the DC SQUID is shown in Fig.~\ref{fig:fig1}(d). The intrinsic capacitance of the Josephson junctions $\rm C_{J1}$ and $\rm C_{J2} $ functions as the parallel capacitor, forming $\rm C_{SQUID}=C_{J1}+C_{J2}$. The equivalent circuit diagram reveals that the DC SQUID in our design is electrically equivalent to a parallel combination of a traditional DC SQUID and a discrete capacitor. The DC bias and the pump tone are combined using a bias tee and subsequently applied to the flux-pump line. The layout of the low-temperature measurement and control circuitry within the dilution refrigerator is presented in the supplementary material.

The circuit of our device is generally consisted of two layers, the base layer and the merged-element Josephson junction (MEJJ) layer. To fabricate the desired circuit, we first deposit a $100 \ \rm nm$ aluminum film on a 2-in c-plane sapphire substrate with an ultrahigh vacuum (UHV) e-beam evaporation system. The desired pattern for the first layer is generated by exposing positive photoresist MICROPOSIT S1813 on the deposited film with a Heidelberg DWL66+ laser lithography system. Then the wafer is developed in the tetramethyl ammonium hydroxide based developer at room temperature. After a wet etching process, we obtain the pattern of the base layer, which defines the impedance transformer circuit, the flux-pump line and the bottom layer of the merged-element Josephson junctions.

In order to obtain the MEJJ layer, we use the bottom undercut layer PMGI SF6 and upper layer MICROPOSIT S1813 to create a bilayer structure. The top electrode pattern of the Josephson junction is defined using a laser lithography system aligned to the base layer. Following the development process, the wafer is loaded into the UHV e-beam evaporation system. An \begin{itshape}in-situ\end{itshape} argon-ion milling process is then employed to remove the native oxide on the bottom electrode of the overlap junctions. Subsequently, the bottom electrode is oxidized in a static oxygen atmosphere at 100-Torr for 1-h to achieve the desired $\rm Al_2O_3$ layer thickness. A 200 nm thick aluminum film is then deposited using an electron beam evaporation system. After liftoff, we obtain the overlap junctions. In our fabrication process, the oxidation conditions are crucial for determining the junction resistance and capacitance. 

\begin{figure}[hptb]
\includegraphics[width=\linewidth]{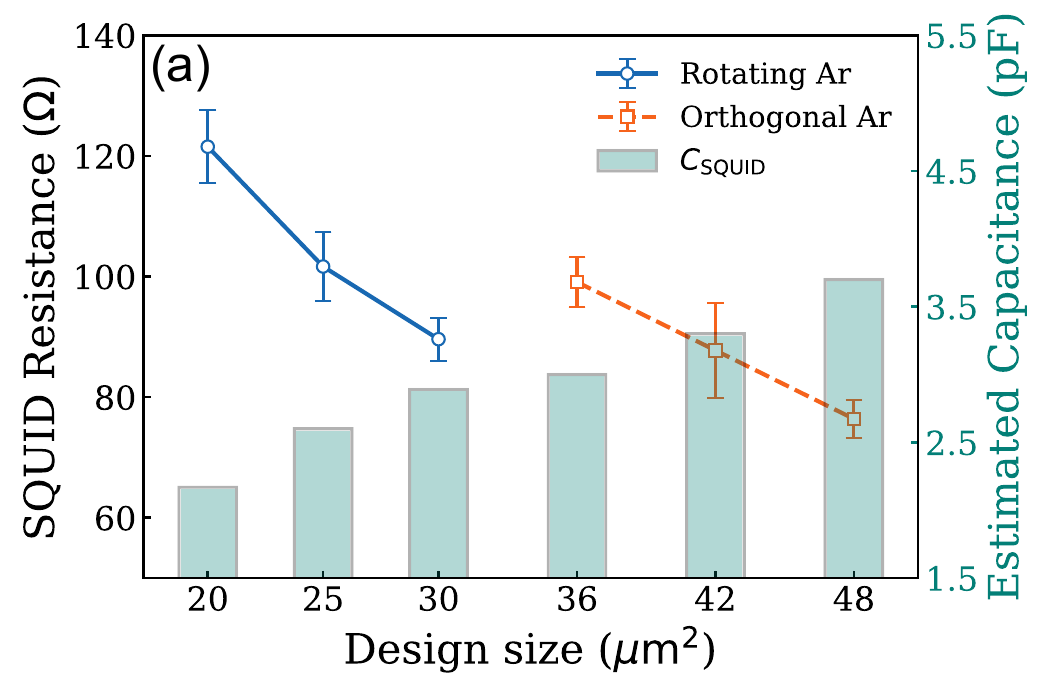}
\includegraphics[width=\linewidth]{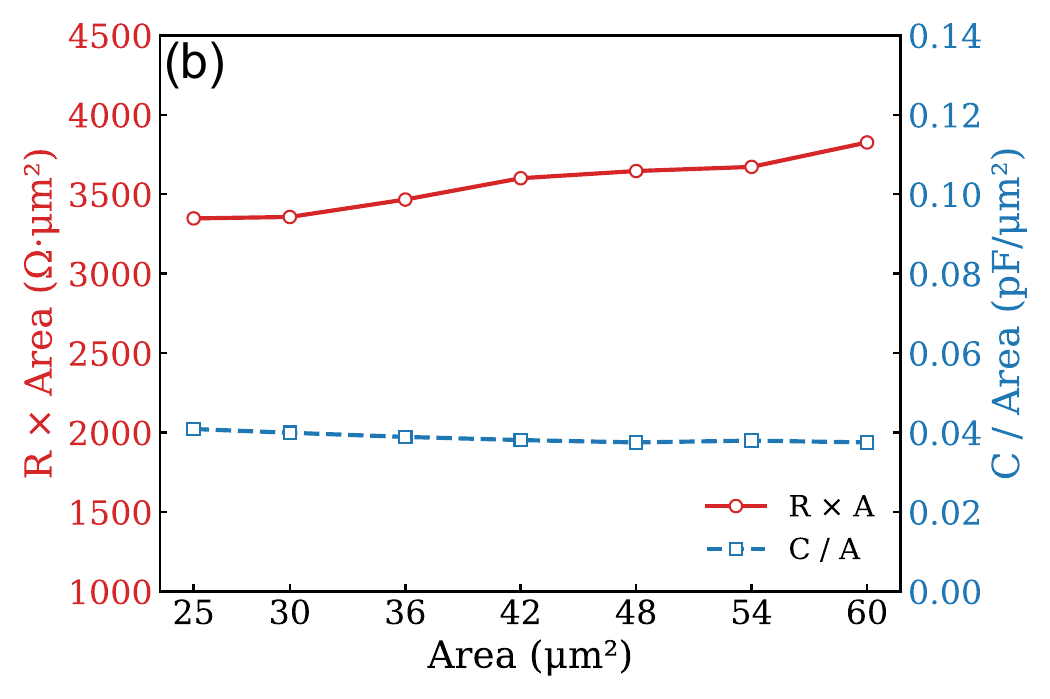}
\caption{\label{fig:2} (a) The measured Room-temperature resistance and capacitance of the SQUIDs for different junction areas. The x-axis represents the defined junction area of a single Josephson junction. The Josephson junctions are fabricated by oxidizing an aluminum film in $100 \ \rm torr$ static oxygen for 1-h. The blue line corresponds to results obtained with chip rotation during argon-ion milling at 10 rpm, while the orange line represents results from continuous argon-ion milling at four orthogonal angles. The teal bars (histogram) show the capacitance estimated from low-temperature resonant frequency measurements. (b)  Resistance-area product (RA, red circles) and  capacitance per unit area (C/A, blue squares) for SQUIDs with different junction sizes. RA values are obtained from room-temperature resistance measurements. C/A values are extracted from low-temperature resonant frequency fitting.}
\end{figure}

To obtain the appropriate MEJPA resonant frequency, we needed to identify the critical current of the SQUIDs. The critical current $\rm I_c$ is calculated using the Ambegaokar-Baratoff formula, $\rm I_c=\pi\Delta/2eR_n$,~\cite{ambegaokarTunnelingSuperconductors1963a} where $\rm \Delta$ is the superconducting energy gap and $\rm R_n$ is the SQUID resistance measured at room temperature. We fabricated and tested the room temperature resistance of SQUIDs with varying junction areas $\rm A_{JJ}$ under oxidation conditions of 100-Torr for 1-h, as shown in Fig.~\ref{fig:2}(a). Considering the large-area nature of the overlap junctions, we tested two different argon-ion milling methods to improve the process stability. One method involved rotating the chip at 10 rpm during milling, while the other performed milling from four orthogonal directions. The error bars reflect the uniformity of junction resistance across the entire 2-inch chip. In this work, we chose the second milling method in the fabrication process. Based on the Ambegaokar-Baratoff relation and our resistance measurements, we estimate a typical critical current density of $J_c \approx 40~\mathrm{nA/\mu m^2}$.
Combining the SQUID room-temperature resistance with the peak resonant frequency measured at low temperatures, we deduced the capacitance value $\rm C_{SQUID}$ using Eq.\ref{eq4}, as represented by the teal bars in Fig.~\ref{fig:2}(a). Based on the results, we can estimate the capacitance density of the SQUID is about $38 \ \rm fF/\mu m^2$. To provide quantitative insight into the junction parameters relevant to device performance, we fabricated MESQUIDs with varying junction areas and characterized their resistance and capacitance. The resistance-area product (RA) and the capacitance per unit area (C/A) are plotted in Fig.~\ref{fig:2}(b). The RA values were directly obtained from room-temperature resistance measurements, while C/A values were deduced from the resonant frequencies of the corresponding nonlinear oscillators at low temperatures. The slight variation in RA values may in part originate from not accounting for the junction sidewall area in the defined junction size~\cite{balOverlapJunctionsSuperconducting2021}. These results validate the consistency of our fabrication process and provide critical metrics for comparing junction uniformity and capacitance density. The specific design parameters of our parametric amplifier are detailed in Table~\ref{tab:label1}. 

\begin{figure}[hptb]
\includegraphics[width=\linewidth]{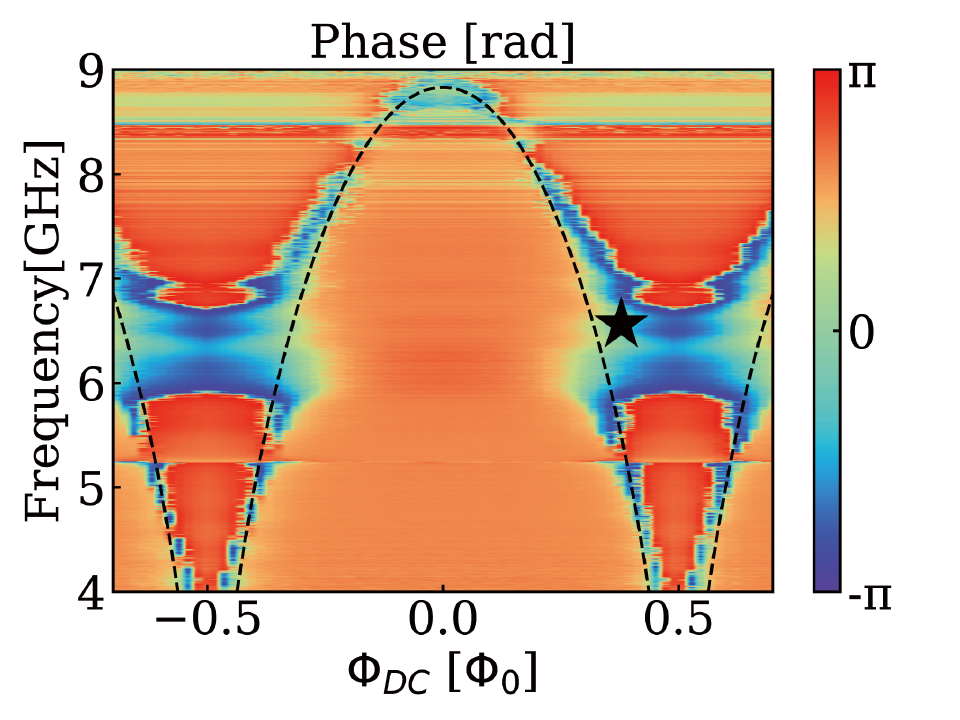}
\caption{\label{fig:fig3}Phases of the transmission spectra $\rm S_{21}$ as a function of DC flux $\rm \Phi_{DC}$ without pump. The resonant frequency of the nonlinear resonator can be tuned from 4 to approximately 9 GHz. Black dashed lines represent the calculated resonant frequencies. The black pentagram indicates the operating point, where $\rm \Phi_{DC}=0.334 \ \Phi_0$.}
\end{figure}

In our experiment, the device operates at three-wave mixing mode with a flux pump frequency of $\omega_p \approx 2\omega_s$. While our device can also operate as a four-wave mixing amplifier, three-wave mixing operation avoids frequency collisions and prevents saturation of the subsequent high-electron-mobility transistor (HEMT) amplifier due to excessive pump power. In the absence of pumping, we measured the transmission spectra $\rm S_{21}$ under different DC magnetic flux biases, as shown in Fig.~\ref{fig:fig3}. Due to the periodic flux dependence of the Josephson inductance in the SQUID, the resonant frequency exhibits a periodic variation with a period of one magnetic flux quantum $\rm \Phi_0$. Figure.~\ref{fig:fig3} presents the resonance tuning over a single flux period, where the frequency varies from approximately 4 GHz to 9 GHz. The device demonstrates repeated operational windows due to this periodicity, and the flux bias point $\rm \Phi_{DC}=0.334 \ \Phi_0$, indicated by the black pentagram, was chosen for optimal performance in our gain and noise measurements. The black dashed line indicates the resonant frequency calculated by the following equation~\cite{mutusDesignCharacterizationLumped2013}:
\begin{equation}\label{eq4}\rm
    \Omega_p(\Phi_{DC})=\frac{1}{\sqrt{C_J(L_{J}/cos(\pi \Phi_{DC}/\Phi_0)+L_s)}}.
\end{equation}

\begin{table}[tbp]
    \centering
    \setlength{\tabcolsep}{2.6mm}
    \begin{tabular}{ccccccc}
        \toprule
        $\omega_t$ & $\rm Z_{\lambda/4}$ & $\rm Z_{\lambda/2}$ & $\rm I_{SQUID}$ & $\rm C_{SQUID}$ & $\rm A_{JJ}$\\
        \hline
        6.5 GHz & $38.5 \ \Omega$ & $65 \ \Omega$ & $3.8 \ \mu A$ & 3.6 pF & $48 \ \mu m^2$ \\
        \hline
    \end{tabular}
    \caption{The design parameters of MEJPA.}
    \label{tab:label1}
\end{table}

\noindent Here, $\rm \Phi_{DC}$ is the dc bias magnetic flux penetrating the SQUID loop. $\rm \Phi_0$ is the flux quantum. $\rm C_J$ and $\rm L_J$ are the capacitance and the Josephson inductance of the SQUID, respectively. $\rm L_s$ is the stray inductance between the SQUID and the ground. As the HEMT amplifier operates within the frequency range of 4-8 GHz, a significant deterioration in the signal-to-noise ratio is observed above 8 GHz in Figure.~\ref{fig:fig3}.

\begin{figure}[hptb]
\includegraphics[width=\linewidth]{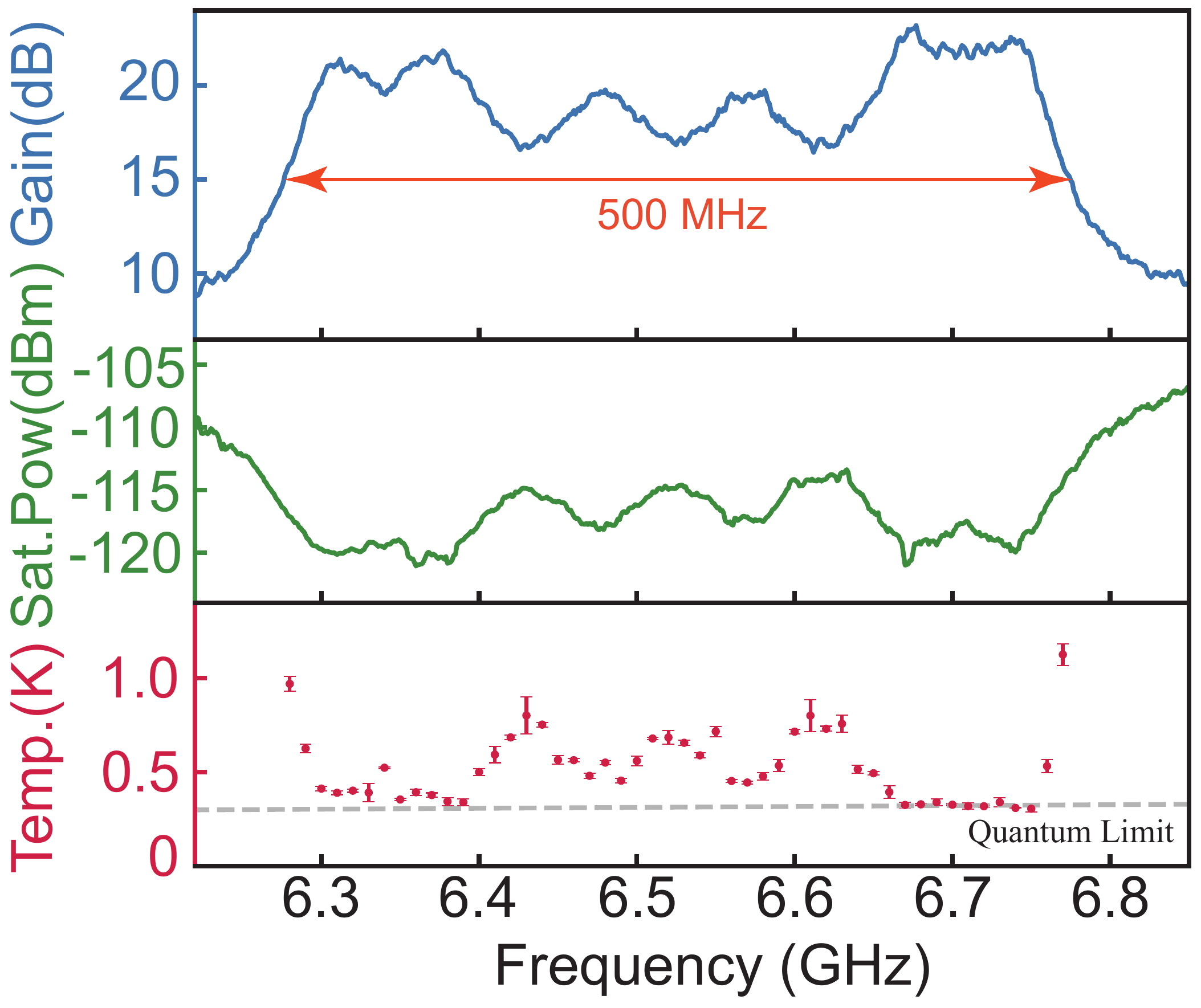}
\caption{\label{fig:fig4}The gain profile (a), saturation power (b), and noise performance (c) of the MEJPA. The operating point is indicated by the black pentagram marker in Fig.~\ref{fig:fig3}, with pump frequency $\rm f_{pump} = 13.04715\ GHz$, pump power $\rm P_{pump}\approx -53\ dBm$, and magnetic flux bias $\rm \Phi_{DC}=0.334\ \Phi_{0}$. The ripple observed across the amplified band originates from standing waves caused by residual impedance mismatch in the measurement chain~\cite{zhou2025impedance,royBroadbandParametricAmplification2015}.} 
\end{figure}

To characterize the amplification performance of our device, we optimized the operating point around 6.5 GHz. A flux pump with an on-chip power of $\rm P_{pump}\approx -53\ dBm$ and a frequency of $\rm f_{pump}=13.04715\ GHz$ was applied to the flux-pump line. This on-chip value was determined by accounting for all cryogenic line attenuations and confirmed via calibration. Our calibration approach, combining standard two-port VNA measurements with independently measured cryogenic line losses, allowed accurate extraction of the intrinsic $\rm S_{21}$ response. In the course of this work, we became aware of a more systematic in-operando calibration methodology proposed for parametric amplifiers~\cite{shin2024operando}, which may offer further improvements for future amplifier characterization. Under these conditions, an instantaneous gain exceeding 15 dB was achieved over a 500 MHz bandwidth, as shown in Fig.~\ref{fig:fig4}(a). Moreover, gain performance exceeding 10 dB was also achieved under other parameters, with the maximum bandwidth surpassing 1 GHz. (see the supplementary material for details). The ripple observed in the gain profile is attributed to the frequency-dependent impedance mismatch caused by standing waves in the input line, such as reflections from the circulator cable or bonding interfaces. These cause periodic variations in the input admittance seen by the amplifier, which in turn modulate the gain, as discussed in Ref.~\cite{royBroadbandParametricAmplification2015}. To qualitatively understand how key design parameters affect amplifier performance, we performed numerical simulations of gain profiles as functions of the MESQUID shunt capacitance and the characteristic impedances of the $\lambda/2$ and $\lambda/4$ transmission line sections. The details of these simulations are also provided in the supplementary material. The MEJPA's linear amplification range was characterized by measuring the 1 dB compression power across its gain bandwidth. By gradually increasing the input signal power, we determined the 1 dB compression point, which corresponds to an average saturation power of approximately -116 dBm [Fig.~\ref{fig:fig4}(b)].

In addition to gain and bandwidth, noise temperature is a crucial performance metric for parametric amplifiers. In this work, we determined the noise temperature $\rm T_{JPA}$ of the MEJPA using the $\rm \Delta SNR$ method,~\cite{Simoenmultimode2015,royBroadbandParametricAmplification2015,duanBroadbandFluxpumpedJosephson2021}
\begin{align}\label{eq5}
    \rm T_{JPA}&=\rm T_{sys}(\frac{1}{\Delta \rm SNR}-\frac{1}{G_{JPA}})+T_{in}(\frac{1}{\Delta SNR}-1), \\
    \rm T_{sys}&=\rm T_{in}+T_{JPA}+T_{HEMT}/G_{JPA}.
\end{align}

\noindent Here, $\rm T_{HEMT}$ is the HEMT noise calibrated by the Y-factor method,~\cite{mutusStrongEnvironmentalCoupling2014a,eloBroadbandLumpedelementJosephson2019,pozar2015microwave} $\rm\Delta SNR$ represents the improvement of signal-to-noise (SNR) between the pump on and off, and $\rm T_{in}$ is the input noise temperature, which is approximately equal to the vacuum noise temperature. The results presented in Fig.~\ref{fig:fig4}(c) demonstrate that the noise of our amplifier approaches the quantum limit.~\cite{cavesQuantumLimitsNoise1982,RevModPhys2010} The circuit diagram of the noise measurement setup can be found in the supplementary material.

In conclusion, we have presented a merged-element technique for designing and fabricating high-performance broadband Josephson parametric amplifiers. Our design eliminates the need for discrete capacitors, simplifying the fabrication process, reducing the requirements for high-precision exposure tools, and ensuring compatibility with standard superconducting qubit fabrication procedures. Furthermore, the increased oxidation layer thickness enhances the dielectric strength of the Josephson junctions, effectively mitigating the electrostatic failure risk of the MEJPA. According to the simulation results, the gain profile can be further optimized by fine-tuning the impedance and control parameters. Our results demonstrate a promising Josephson parametric amplifier suitable for multi-qubit measurement architectures and with broader implications for quantum information processing.

See the supplementary material for detailed descriptions of the experimental setup, additional gain-bandwidth results, numerical simulations illustrating design parameter dependencies, and performance benchmarking of the MEJPA against other IMPAs utilizing discrete capacitor implementations.

This work was partly supported by the Innovation Program for Quantum Science and Technology (2021ZD0301702), the National Natural Science Foundation of China (NSFC) (U21A20436, 12074179 and 12205147), the Natural Science Foundation of Jiangsu Province, China (Grant Nos. BE2021015-1 and BK20232002), the Natural Science Foundation of Shan-dong Province (Grant No. ZR2023LZH002).


\section*{Data Availability}
The data that support the findings of this study are available from the corresponding author upon reasonable request.

\section*{REFERENCES}
\bibliography{MEJPA}

\newpage

\end{document}


\title{Supplementary Materials for \\
    Broadband merged-element Josephson parametric amplifier}

\author{Yuting Sun}
\email[]{qu_syt@163.com}
\affiliation{National Laboratory of Solid State Microstructures, School of Physics, Nanjing University, Nanjing 210093, China}
\affiliation{Shishan Laboratory, Suzhou Campus of Nanjing University, Suzhou 215163, China}
\affiliation{Hefei National Laboratory, Hefei 230088, China}
\author{Xianke Li}
\affiliation{National Laboratory of Solid State Microstructures, School of Physics, Nanjing University, Nanjing 210093, China}
\affiliation{Shishan Laboratory, Suzhou Campus of Nanjing University, Suzhou 215163, China}
\affiliation{Hefei National Laboratory, Hefei 230088, China}
\author{Qingyu Wang}
\affiliation{National Laboratory of Solid State Microstructures, School of Physics, Nanjing University, Nanjing 210093, China}
\affiliation{Shishan Laboratory, Suzhou Campus of Nanjing University, Suzhou 215163, China}
\author{Tairong Bai}
\affiliation{National Laboratory of Solid State Microstructures, School of Physics, Nanjing University, Nanjing 210093, China}
\affiliation{Shishan Laboratory, Suzhou Campus of Nanjing University, Suzhou 215163, China}
\affiliation{Hefei National Laboratory, Hefei 230088, China}
\author{Xudong Liao}
\affiliation{National Laboratory of Solid State Microstructures, School of Physics, Nanjing University, Nanjing 210093, China}
\affiliation{Shishan Laboratory, Suzhou Campus of Nanjing University, Suzhou 215163, China}
\author{Dong Lan}
\affiliation{National Laboratory of Solid State Microstructures, School of Physics, Nanjing University, Nanjing 210093, China}
\affiliation{Shishan Laboratory, Suzhou Campus of Nanjing University, Suzhou 215163, China}
\affiliation{Hefei National Laboratory, Hefei 230088, China}
\affiliation{Synergetic Innovation Center of Quantum Information and Quantum Physics, University of Science and Technology of China, Hefei, Anhui 230026, China}
\author{Jie Zhao}
\email[]{jiezhao@nju.edu.cn}
\affiliation{National Laboratory of Solid State Microstructures, School of Physics, Nanjing University, Nanjing 210093, China}
\affiliation{Shishan Laboratory, Suzhou Campus of Nanjing University, Suzhou 215163, China}
\affiliation{Hefei National Laboratory, Hefei 230088, China}
\author{Yang Yu}
\affiliation{National Laboratory of Solid State Microstructures, School of Physics, Nanjing University, Nanjing 210093, China}
\affiliation{Shishan Laboratory, Suzhou Campus of Nanjing University, Suzhou 215163, China}
\affiliation{Hefei National Laboratory, Hefei 230088, China}
\affiliation{Synergetic Innovation Center of Quantum Information and Quantum Physics, University of Science and Technology of China, Hefei, Anhui 230026, China}

\maketitle

\begin{center}
    \textbf{EXPERIMENTAL SETUP}
\end{center}

The simplified schematic of the dilution refrigerator used to characterize the MEJPA is shown in Fig.~\ref{fig:setup}. In order to switch different noise source lines, we used two Radiall low-temperature microwave switches to construct noise measurement system.

During the calibration of the amplifier, the network analyzer generates an input signal at frequency $\omega_s$, which undergoes successive attenuation before reaching the mixing chamber (MC) plate. The output signal, reflected from the sample, is then routed back to the network analyzer via circulators, isolators, a HEMT (high-electron-mobility transistor), and a room-temperature amplifier. In the experiment, a three-wave mixing technique was used to drive the devices through the flux modulation port. The dc signal and the microwave pump signal were combined at room temperature using a bias tee and then applied to the fast-bias line.

\renewcommand{\thefigure}{S\arabic{figure}}
\begin{figure}[hptb]
    \centering
    \includegraphics[width=0.56\linewidth]{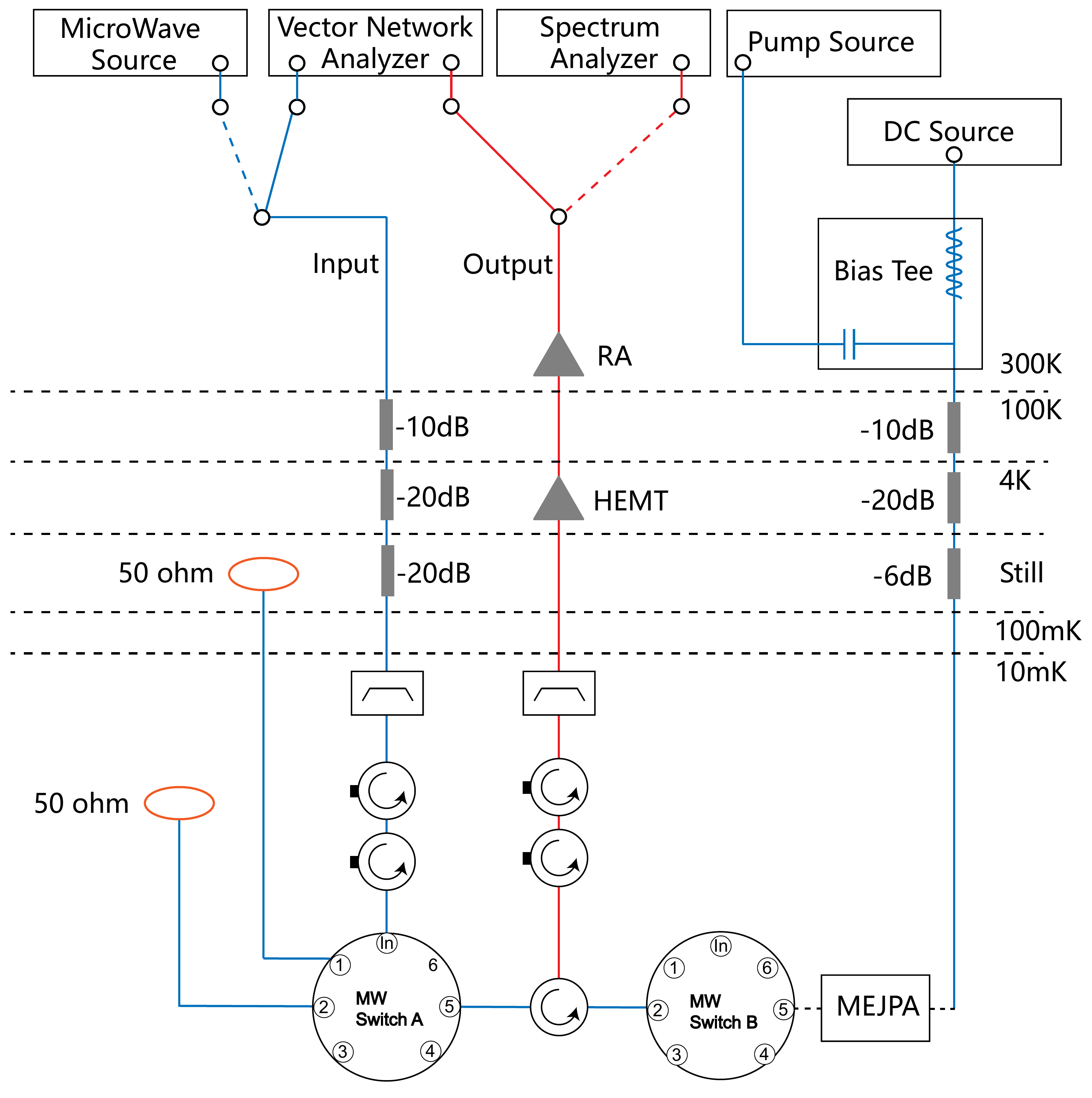}
    \caption{Simplified schematic of the characterization of the MEJPA. Two 50 $\Omega$ loads were used as noise sources. The hot source wat heat sunk to the 1K stage of the dilution fridge with a base temperature of 1.1K, and the cold noise source was placed on the mixing chamber with a base temperature of $\rm \approx 10mK$. Two single-pole, six-throw cryogenic microwave switches are used to switch measurement lines.}
    \label{fig:setup}
\end{figure}

To determine the intrinsic $\rm S_{21}$ response of the MEJPA, we employed a two-port VNA calibration (through-open-load) and independently measured the cryogenic line attenuation using calibrated signal sources and known cable losses at each temperature stage. This allowed accurate estimation of the on-chip signal and pump power. 

To measure the amplifier noise, an input signal is generated by a microwave source and received by a spectrum analyzer with the pump both on and off. The system noise of the output line was characterized using the Y-factor method, which employed a cryogenic microwave switch to alternate between a 50-ohm termination at the still plate (acting as the hot source) and the MC stage (acting as the cold source).

\begin{center}
    \textbf{OTHER RESULTS OF MEJPA}
\end{center}

In the low-temperature characterization, we also measured the amplifier bandwidth corresponding to a 10~dB gain under varying magnetic flux bias conditions. As shown in Fig.~\ref{fig:gain234}, the bandwidth exceeding 10~dB reached up to 1~GHz. However, the gain flatness remains suboptimal and could be further improved through enhanced impedance matching.

\renewcommand{\thefigure}{S\arabic{figure}}
\begin{figure}[hptb]
    \centering
    \includegraphics[width=0.45\linewidth]{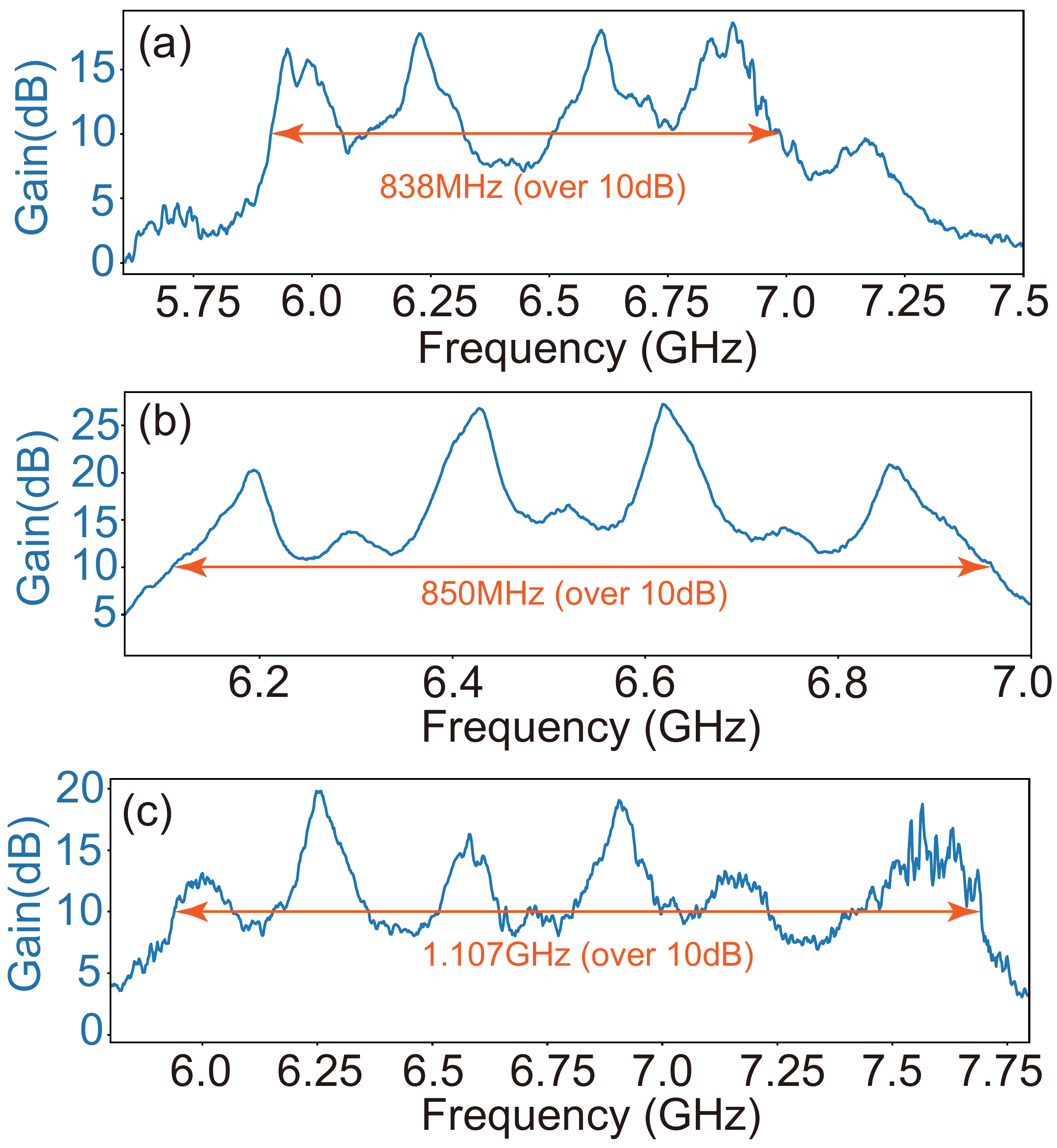}
    \caption{The 10 dB gain profiles (a)(b)(c) of different devices are shown, with the orange line indicating the bandwidth where the gain exceeds 10 dB.}
    \label{fig:gain234}
\end{figure}

\begin{center}
    \textbf{NUMERICAL SIMULATIONS OF DESIGN PARAMETER DEPENDENCIES}
\end{center}

To illustrate how key structural parameters in the MEJPA design influence amplifier performance, we performed numerical simulations based on a nonlinear resonator model comprising a Josephson junction and embedded transmission lines~\cite{royBroadbandParametricAmplification2015}. The gain response was calculated under variations of three critical parameters.

Figure~\ref{fig:simulation}(a) shows the dependence of gain bandwidth and peak gain on the MESQUID shunt capacitance $C_s$, which is integrated directly into the SQUID structure. Figures~\ref{fig:simulation}(b) and (c) present the effects of the characteristic impedances $Z_{\lambda/2}$ and $Z_{\lambda/4}$ of the half- and quarter-wavelength coplanar waveguide sections, respectively. These two sections together control the impedance transformation and influence the peak gain, ripple behavior, and bandwidth matching.

These simulation results provide guidance for optimizing design parameters to achieve broadband and flat gain performance in merged-element architectures.

\renewcommand{\thefigure}{S\arabic{figure}}
\begin{figure}[hptb]
    \centering
    \subcaptionbox{}
    {\includegraphics[width=0.32\linewidth]{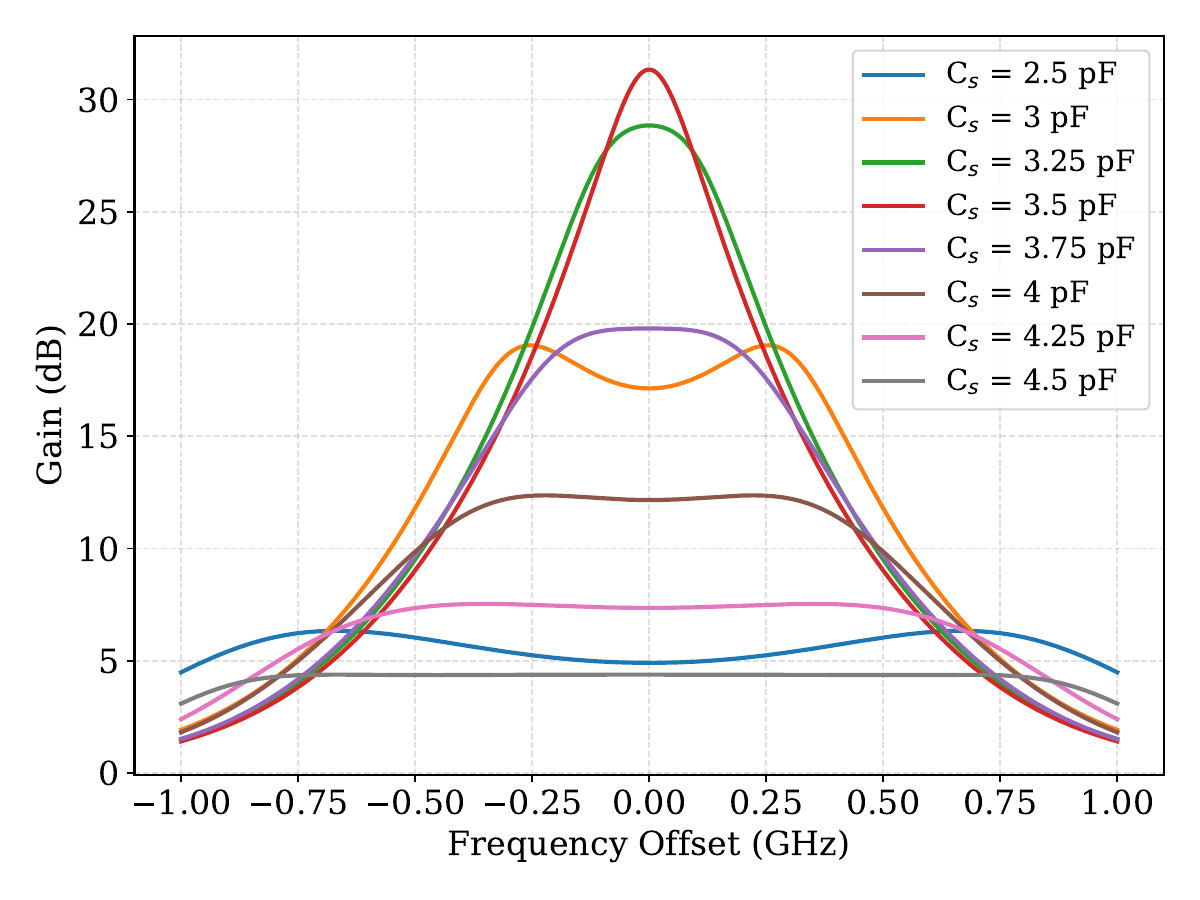}}
    \subcaptionbox{}
    {\includegraphics[width=0.33\linewidth]{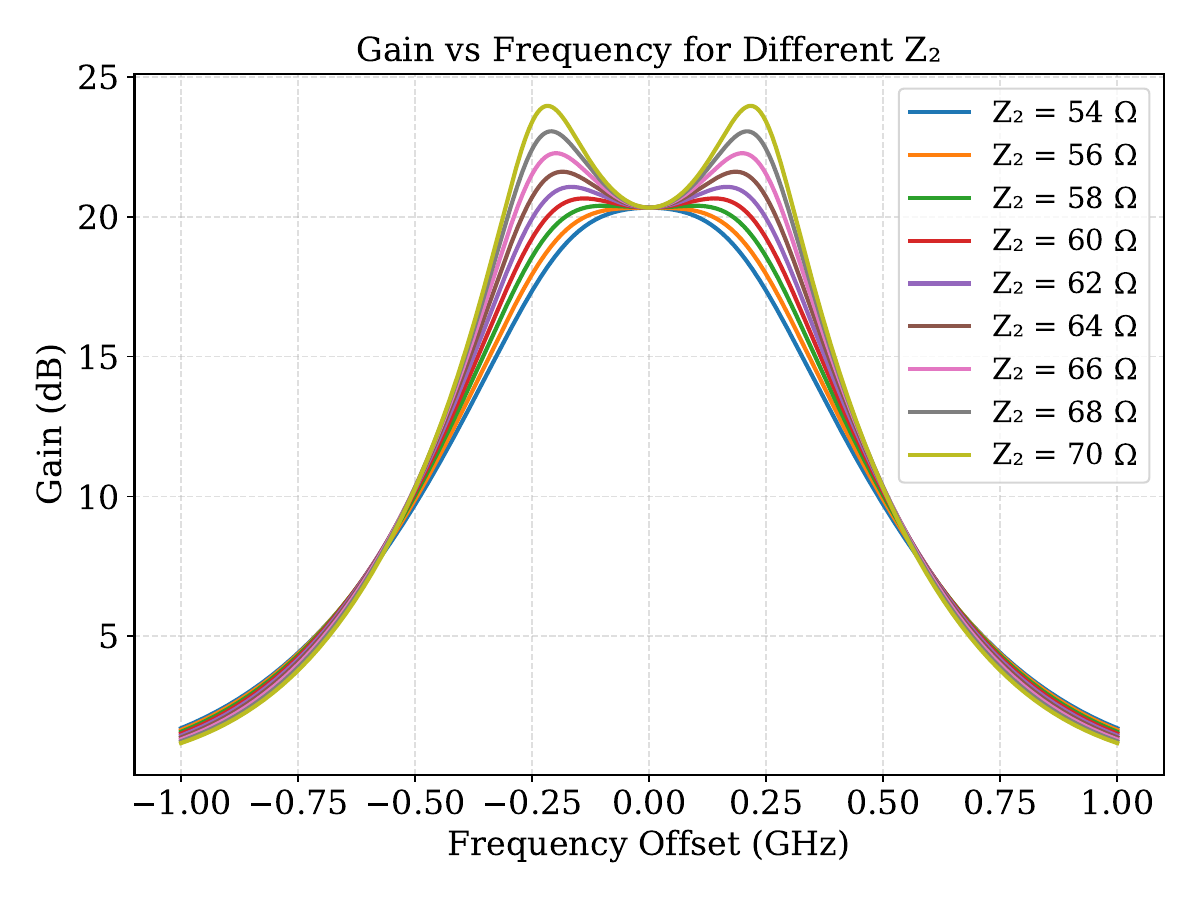}}
    \subcaptionbox{}
    {\includegraphics[width=0.33\linewidth]{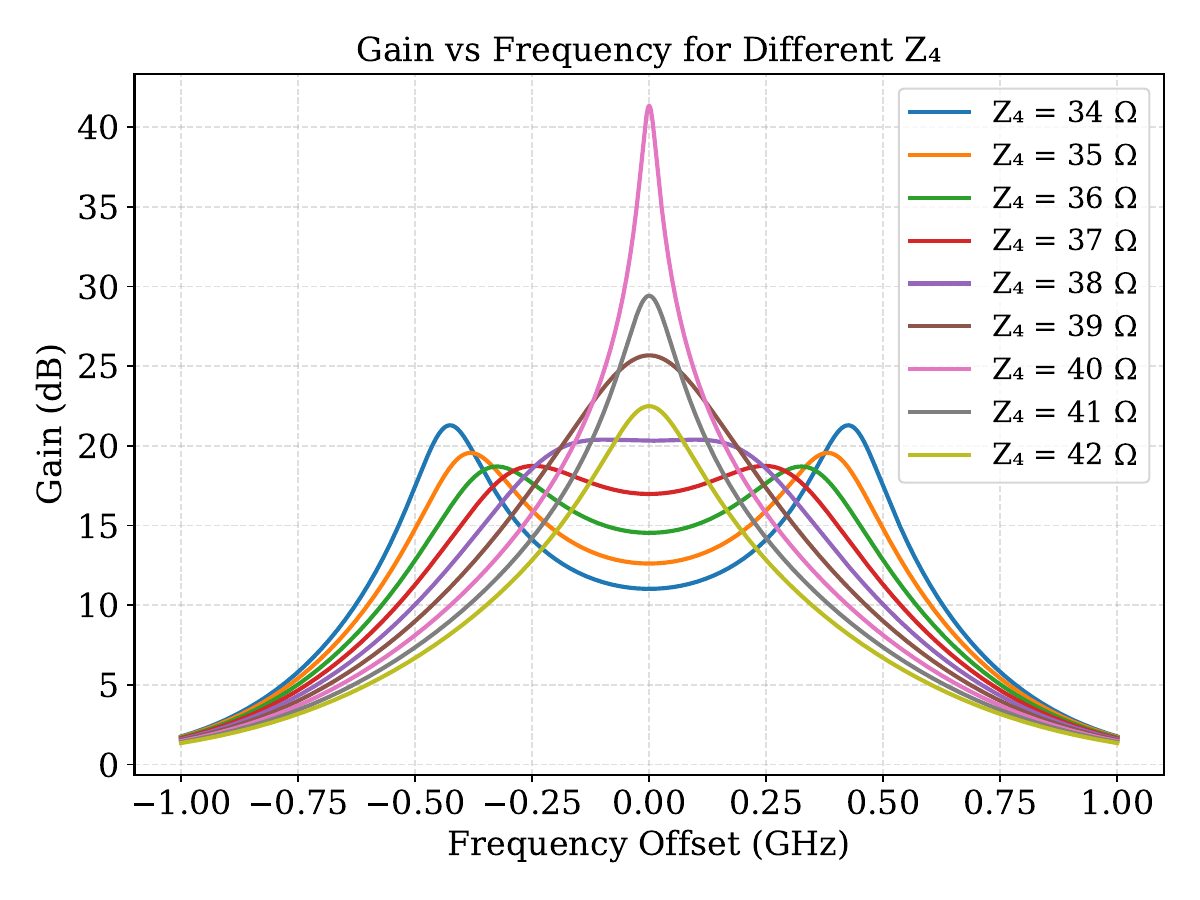}}
    \caption{Simulated gain profiles showing dependence on key structural parameters: (a) MESQUID shunt capacitance $C_s$; (b) Characteristic impedance $Z_{\lambda/2}$ of the half-wavelength section; (c) Characteristic impedance $Z_{\lambda/4}$ of the quarter-wavelength matching section.}
    \label{fig:simulation}
\end{figure}

\newpage
\begin{center}
    \textbf{IMPA PERFORMANCE COMPARISON WITH DIFFERENT CAPACITOR DESIGNS}
\end{center}

To highlight the advantages of merging the shunting capacitor into the SQUID structure, and coupling it via the impedance transformer. We compare our MEJPA design with prior broadband JPAs that employ separate lumped capacitors. Table~S1 summarizes key performance parameters, including gain, bandwidth, saturation power, on-chip pump power, and qualitative fabrication complexity.

\vspace{1em}

\renewcommand{\thetable}{S\arabic{table}}
\begin{table}[hptb]
\centering
\caption{\centering{Benchmarking of the MEJPA against other broadband JPAs.}}
\begin{tabular}{lccccc}
\hline
\textbf{Device} & \textbf{Gain (dB)} & \textbf{BW (MHz)} & \textbf{$P_{1\mathrm{dB}}$ (dBm)} & \textbf{Pump Power (dBm)} & \textbf{Fabrication Complexity} \\
\hline
\textbf{This work (MEJPA)} & 15 & 500 & $-116$ & $-53$ & \textbf{Low} (merged capacitor) \\
Duan et al. (2021)\cite{duanBroadbandFluxpumpedJosephson2021} & 15 & 600 & $-110$ & $-55^{\dagger}$ & Moderate (separate dielectric $C_p$) \\
Wu et al. (2022)\cite{wuVacuumgapbasedLumpedElement2022} & 20 & 162 & $-123$ & $-$ & High (vacuum-gap-based $C_p$) \\
\hline
\end{tabular}

\textit{$\dagger$Estimated from on-chip attenuation and pump line loss, based on supplementary information. The third row does not report on-chip pump power.}
\end{table}

\vspace{1em}

Compared to parametric amplifier designs with discrete capacitors, the MEJPA integrates the capacitive element directly into the SQUID structure, eliminating the need for separate capacitor components. This merged-element configuration significantly reduces parasitic inductance, improves impedance matching, and simplifies the fabrication process.

\section*{REFERENCES}
\bibliography{MEJPAsupp}